\documentclass[aps,prl,superscriptaddress,twocolumn,showpacs,floatfix]{revtex4-1}
\usepackage{graphicx}
\usepackage{latexsym}
\usepackage{fancybox}
\usepackage{color,soul}

\usepackage{pdfpages}
\usepackage{pgffor}
\makeatletter
\AtBeginDocument{\let\LS@rot\@undefined}
\makeatother

\begin{document}
\title{Signature of frustrated moments in quantum critical CePd$_{1-x}$Ni$_x$Al}
\author{Akito Sakai}
\affiliation{Experimental Physics VI, Center for Electronic Correlations and Magnetism, Institute of Physics, University of Augsburg, 86135 Augsburg, Germany}
\author{Stefan Lucas}
\affiliation{Max Planck Institute for Chemical Physics of Solids, 01187 Dresden, Germany}
\author{Philipp Gegenwart}
\affiliation{Experimental Physics VI, Center for Electronic Correlations and Magnetism, Institute of Physics, University of Augsburg, 86135 Augsburg, Germany}
\author{Oliver~Stockert}
\affiliation{Max Planck Institute for Chemical Physics of Solids, 01187 Dresden, Germany}
\author{Hilbert v. L\"ohneysen}
\affiliation{Karlsruhe Institute of Technology, Institute for Solid State Physics and Physics Institute, 76131 Karlsruhe, Germany}
\author{Veronika Fritsch}
\affiliation{Experimental Physics VI, Center for Electronic Correlations and Magnetism, Institute of Physics, University of Augsburg, 86135 Augsburg, Germany}

\date{\today}

\begin{abstract}
CePdAl with Ce $4f$ moments forming a distorted kagom\'{e} network is one of the scarce materials exhibiting Kondo physics and magnetic frustration simultaneously. As a result, antiferromagnetic (AF) order setting in at $T_{\mathrm{N}} = 2.7$~K encompasses  only two thirds of the Ce moments. We report measurements of the specific heat, $C$, and the magnetic Gr\"uneisen parameter, $\Gamma_{\rm mag}$, on single crystals of CePd$_{1-x}$Ni$_x$Al with $x\leq 0.16$ at temperatures down to 0.05~K and magnetic fields $B$ up to $8$~T. Field-induced quantum criticality for various concentrations is observed with the critical field decreasing to zero at $x_c\approx 0.15$.
 Remarkably, two-dimensional (2D) AF quantum criticality of Hertz-Millis-Moriya type arises for $x=0.05$ and $x=0.1$ at the suppression of 3D magnetic order. Furthermore, $\Gamma_{\rm mag}(B)$ shows an additional contribution near $2.5$~T for all concentrations which is ascribed to correlations of the frustrated one third of Ce moments.

\end{abstract}

\maketitle

New quantum states of matter, arising in materials with competing interactions and multiple energetically degenerate configurations are of strong interest in condensed matter physics. For example, unconventional superconductivity is found near quantum critical points (QCPs)~\cite{Mathur1998Magnetically,Ramshaw2015Quasiparticle,Steglich1979Superconductivity,Sachdev2010Where,Stewart2011Superconductivity} and spin-liquid phases, driven by strong frustration have been realized in magnetic insulators~\cite{Ramirez1994Strongly,Balents2010Spin}. However, there are only few studies on metallic frustrated magnets~\cite{Fritsch2006Correlation}, and the effect of frustration on quantum criticality in metals has rarely been investigated experimentally. Rare-earth heavy-fermion (HF) metals, consisting of $4f$ moments coupled to conduction electrons by an exchange interaction $J$, are ideally suited for this purpose. Since $J$ governs the competition between the indirect Ruderman-Kittel-Kasuya-Yosida (RKKY) exchange and the Kondo interaction~\cite{Doniach1977Kondo}, QCPs can be realized by variation of $J$ with pressure, chemical composition or magnetic field~\cite{Stewart2001Non,Loehneysen2007Fermi,Gegenwart2008Quantum}.
Unconventional quantum criticality with a Kondo breakdown and a spin-liquid phase of localized $4f$ moments being decoupled from conduction electrons has been predicted for high degree of frustration \cite{Senthil2004Weak}. The `global phase diagram', which classifies the electronic and magnetic ground states for HF systems, treats $J$ and the strength of quantum fluctuations arising from frustration as two independent parameters~\cite{Si2006Global,Vojta2008From,Coleman2010Frustration}.

The effect of geometrical frustration in Kondo lattices has been recognized in hexagonal systems crystallizing in the ZrNiAl structure. Here, the $4f$ electrons form a structure of equilateral corner-sharing triangles in the $ab$ plane, which can be described as a distorted kagom\'{e} network. For YbAgGe, the geometrical frustration leads to a series of almost degenerate magnetic states tuned by magnetic fields and novel quantum bicritical behavior \cite{Budko2004Magnetic,Tokiwa2013Quantum,Dong2013Anomalous}. CeRhSn does not display long-range order and is located close to a QCP driven by geometrical frustration~\cite{Tokiwa2015Characteristic}. In CePdAl the magnetic frustration gives rise to unusual magnetic ordering \cite{Doenni1996Geometrically,Oyamada1996NMR,Lacroix2010Frustrated,Motome2010Partial}. It has been shown previously using polycrystals  that the material can be tuned through a QCP with the aid of chemical pressure, realized by partial substitution of Pd by Ni~\cite{Fritsch2014Approaching}.

CePdAl exhibits a strong magnetic anisotropy with the susceptibility ratio $\chi_c/\chi_{ab} \approx 15$ at $T = 5$~K, which is attributed to crystalline-electric-field and exchange anisotropies \cite{Isikawa1996Magnetocrystalline}.
Upon cooling, the electrical resistivity exhibits a $-\ln T$ dependence from 20 to 6~K, followed by a coherence maximum at $4$~K and a subsequent decrease \cite{Kitazawa1994Electronic,Fritsch2015Role}.  The presence of the Kondo effect in CePdAl is confirmed by a maximum of the thermopower at 8~K \cite{Huo2002Thermoelectric}. Below $T_{\rm N} = 2.7$ K partial AF order is observed \cite{Doenni1996Geometrically}: despite being crystallographically equivalent, there appear three magnetically inequivalent  sites.  Ce(1) and Ce(3) exhibit long-range ordered moments, while Ce(2) do not, even down to lowest $T$~\cite{Oyamada2008Ordering}. The ordered Ce moments form ferromagnetic chains in the basal plane, which are coupled antiferromagnetically and separated by the latter~\cite{Doenni1996Geometrically,Nunez-Regueiro1997Magnetic}. Along the crystallographic $c$-axis an incommensurate sinusoidal modulation of the magnitude of the ordered moments indicates an overall three-dimensional (3D) magnetic structure of the ordered Ce moments \cite{Doenni1996Geometrically}.  The reduced entropy of less than $0.5 R\ln 2$ at $T_{\rm N}$ \cite{Fritsch2016CePdAl} indicates that all Ce moments are subject to the onset of Kondo screening. Since all Ce sites have similar local environment, it is however unlikely that the Ce(2) moments are completely Kondo screened already at $T_{\rm N}$. Thus, magnetic frustration prevents their long-range order.

We report a detailed thermodynamic study of quantum criticality in single-crystalline CePd$_{1-x}$Ni$_x$Al ($x\leq 0.16$) down to $50$~mK for fields $B$ up to $8$~T applied along the $c$-axis. Besides heat capacity, $C(T,B)$, we focus on the magnetic Gr\"{u}neisen parameter $\Gamma_{\rm mag} = -(dM/dT)_B/C=T^{-1}(dT/dB)_S$ ($M$: magnetization, $S$: entropy). Hence, $\Gamma_{\rm mag}$ can either be calculated from $T$ dependencies of $M$ and $C$ \cite{Tokiwa2009Divergence} or obtained directly as adiabatic magnetocaloric effect~\cite{Tokiwa2011High}. At any field-tunable QCP, $\Gamma_{\rm mag}$ displays a universal divergence and sign change upon tuning the field across the critical field~\cite{Zhu2003Universally,Garst2005Sign}. Our thermodynamic investigation of CePd$_{1-x}$Ni$_x$Al reveals 2D field-induced quantum criticality for $x=0.05$ and $x=0.1$ and a (zero-field) concentration tuned QCP at $x_c \approx 0.15$ in accordance with previous results on polycrystals~\cite{Fritsch2014Approaching}. Most interestingly, we find a positive contribution in $\Gamma_{\rm mag}(B)$ near 2.5~T, in addition to quantum critical behavior, indicative of a field-induced suppression of magnetic entropy. Since this feature is observed even beyond $x_c$ and thus cannot be related to the QCP, we associate it with the frustrated Ce(2) moments.

Single crystals of CePd$_{1-x}$Ni$_x$Al were grown by the Czochralski method. The Ni content was determined by atomic absorption spectroscopy (AAS). The homogeneity of the Ni content of the samples was checked by energy dispersive x-ray spectroscopy (EDX) with a spatial resolution of $1\mu$m and found to be better (for $x\leq 0.1$) or comparable (for $x=0.14$) to 1 at$\%$ Ni. The samples were investigated as cast, since annealing causes a structural phase transition \cite{Gribanov2006New,Fritsch2013Magnetization}.
The samples were oriented by Laue back-scattering x-ray diffraction. The magnetic field was always applied along the crystallographic $c$-axis.
The specific heat was measured by the relaxation method in a physical property measurement system at $T >$ 0.4 K and by both the relaxation and the quasi-adiabatic heat-pulse technique in a dilution refrigerator. The adiabatic magnetocaloric effect was measured using a high-resolution alternating-field technique \cite{Tokiwa2011High}.

\begin{figure}[t]
\begin{center}
\includegraphics[width=0.85\linewidth]{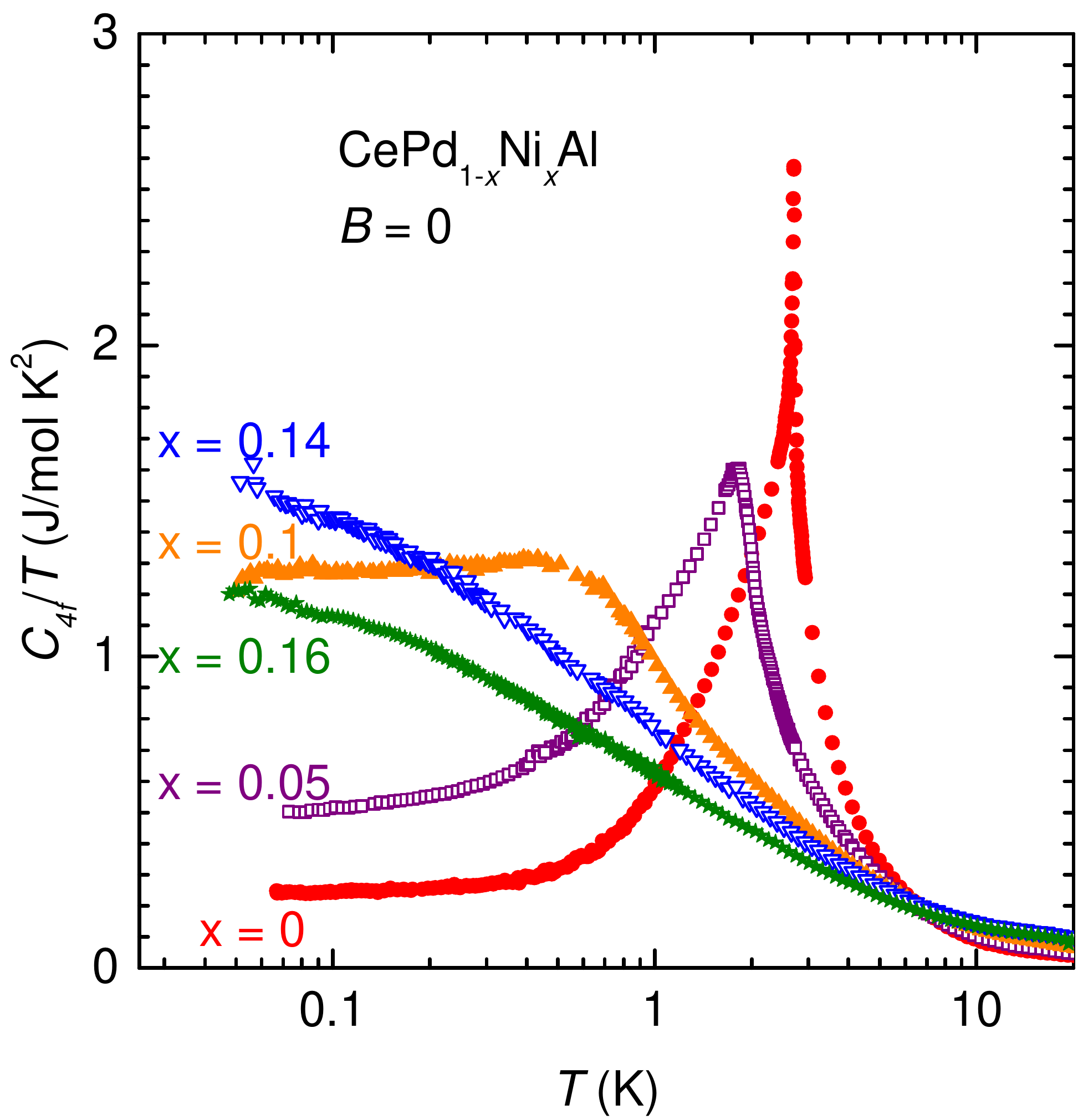}
\caption{4$f$-electron part of the specific heat plotted as $C_{4f}/T$ vs. $T$ (on log scale) for  CePd$_{1-x}$Ni$_x$Al single crystals at zero magnetic field. $C_{4f}$ is obtained after subtracting the nuclear and phonon contributions (see text for details).  \label{fig1}}
\end{center}
\end{figure}

Figure~\ref{fig1} shows the 4$f$-electron contribution $C_{4f}$ to the heat capacity for all investigated single crystals. $C_{4f}$ was obtained after subtraction of a nuclear contribution  $C_{\rm nuc} = A_{\rm n}(B)/T^2$  and phonon and conduction-electron contributions approximated by $C(T)$ of the isostructural nonmagnetic reference compound LuPdAl (see supplemental material (SM) \cite{Supplement} for details).  A sharp peak due to the partial AF order at $T_{\rm N} = 2.7$ K for $x=0$ is suppressed with increasing $x$ to $T_{\rm N} = 1.9$ K for $x=0.05$, as determined by the peak temperatures of $C_{4f}/T$. No long-range order is detectable for $x=0.14$ and $0.16$, again in accordance with the phase diagram obtained on polycrystals \cite{Fritsch2014Approaching}. However, while a logarithmic increase of $C/T$ upon cooling down to 0.06~K has been found for a polycrystal with $x= 0.144$~\cite{Fritsch2014Approaching}, the data for $x=0.14$  clearly display curvature, which may be related to residual order, corroborated by the finite critical field extracted from the field-dependence of the specific heat (see below).

\begin{figure}[t]
\begin{center}
\includegraphics[width=\linewidth]{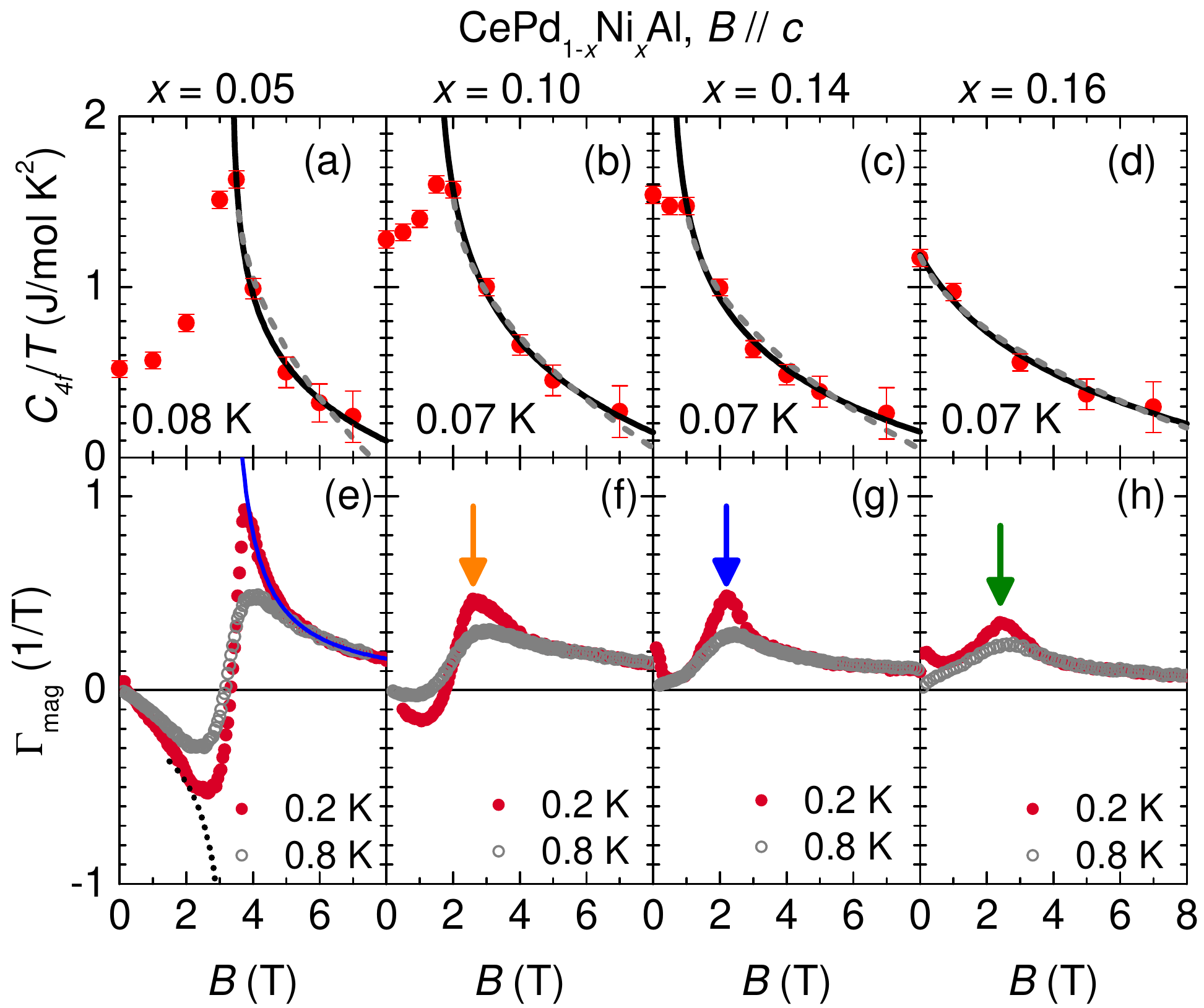}
\caption{(a) - (d): Field dependence ($B\parallel c$) of $C_{4f}/T$ of CePd$_{1-x}$Ni$_x$Al for  $x=0.05$, $0.10$, $0.14$, and $0.16$ at the lowest temperatures. The black thick and gray thin lines represent fits of the data according to the prediction for AF quantum criticality in 2D and 3D, respectively,  using parameters listed in Table \ref{table1}. (e) - (h): Field-dependence of the magnetic Gr\"uneisen parameter, $\Gamma_{\rm mag}$, at $T= 0.2$ and $0.8$~K of CePd$_{1-x}$Ni$_x$Al for $x = 0.05$, $0.10$, $0.14$, and $0.16$. The blue solid line in panel (e) indicates a fit  of the magnetic Gr\"uneisen parameter according to $\Gamma_{\rm mag} = G_r (B - B_c)^{-1}$ with $B_c=3.02 \pm 0.02$~T and $G_r=0.81 \pm 0.01$.  The black dotted line in panel (e) has been obtained by assuming point inversion symmetry of the fit near the critical field.  The arrows in panels (f)-(h) indicate the positions of broad maxima.}\label{fig2}
\end{center}
\end{figure}

\begin{table}[t]
	\caption{ Fit parameters for the curves shown in Fig.~\ref{fig2}(a)-(d). The following functions were fitted~\cite{Zhu2003Universally}: (a) 2D AF quantum criticality according to $C_{4f}/T= -a_1\ln {(a_2 b)}$, $b=(B-B_c)/B_c$, (b) 3D AF quantum criticality according to $C_{4f}/T= \gamma_0- c \sqrt{B-B_c}$. For the discussion of the error bars see SM \cite{Supplement}.}
	\begin{ruledtabular}
\begin{tabular}{lccccc}
                          &    & $x=0.05$	& $x=0.10$	&  $x=0.14$	   & $x=0.16$	\\
\hline
(a) &    $B_{\rm c}$ (T)        &$3.4$	    &$1.5 $     &$0.52$        &$-1.55 $	\\
&	 $a_1$ (J/molK$^2$)         &$0.43$     &$0.56$       &$0.48$	    &$0.55$	\\
  &	$a_2$                       &$0.58$     &$0.18$       &$ 0.052$	    &$ -0.11 $	\\
\hline
(b) &    $B_{\rm c}$ (T)        &$ 3.5$     & $2.0$        &  $0.98$     &  $-0.23$ 	\\
&	 $\gamma_0$ (J/molK$^2$)    &$  1.6$    &  $1.6$       & $1.5$      & $1.4$	     \\
&  	$c$ (J/molK$^2$)            &$ 0.81 $    & $0.63$       &  $0.56$	&   $0.42$  \\
\end{tabular}
	\end{ruledtabular} \label{table1}
\end{table}

In Figure~\ref{fig2}~(a)-(d) the field dependence of $C_{4f}/T$ of CePd$_{1-x}$Ni$_x$Al single crystals at the lowest attainable temperature, obtained from $T$-dependent measurements in constant field shown in the Supplemental Material (SM) \cite{Supplement}, is analyzed.
Assuming a field-induced QCP, $C_{4f}/T= -a_1\ln {(a_2 b)}$ using $b=(B-B_c)/B_c$ and $C_{4f}/T = \gamma_0 -c \sqrt{B - B_c}$ are predicted by scaling theory for 2D and 3D antiferromagnetic quantum criticality, respectively \cite{Zhu2003Universally}. Note that in the 3D case, the critical contribution is subtracted from a constant $\gamma_0$ which equals the saturation value at the QCP. Therefore the fitted $\gamma_0$ must not be smaller than the largest measured $C_{4f}/T$ value. Both equations have three free parameters which allows an unbiased comparison of the quality of the fits. The obtained fit parameters are listed in Table~\ref{table1} and the fits are indicated in Fig.~\ref{fig2}~(a) - (d).
Clearly, for $x=0.05$ the quality of the 2D fit is much better compared to its 3D counterpart. On the other hand, with increasing $x$ a clear distinction between 2D and 3D behavior becomes more difficult and finally impossible for $x = 0.16$. Fits the $x=0.16$ data result  in (physically meaningless) negative critical fields, indicating that this sample lies already  beyond the concentration-tuned QCP as   noted above. Analysis of the field dependence of $C_{4f}$ thus indicates 2D quantum critical fluctuations for $x = 0.05$ while more isotropic (3D) criticality at larger $x$, possibly related to the enhanced disorder in the Ni-substituted samples, cannot be excluded. Also, it is under debate whether a 2D-SDW scenario holds exactly at a QCP or some dimensional cross-over always occurs at sufficiently low temperatures \cite{Garst2008Dimensional}.

To further explore the nature of quantum criticality in CePd$_{1-x}$Ni$_x$Al, we consider the magnetic Gr\"uneisen parameter $\Gamma_{\rm mag}$. In the vicinity of any generic field-induced QCP, at low temperatures, a $1/(B-B_c)$ divergence with universal prefactor, as well as, a sign change near the critical field $B_c$, arising from the accumulation of entropy, are expected~\cite{Garst2005Sign}. Figures~\ref{fig2}~(e) - (h) show $\Gamma_{\rm mag}(B)$ at $T= 0.2$ and $0.8$ K for our samples. For $x = 0.05$ and $x=0.1$ a zero-crossing is observed at $B = 3.35$~T and 1.8~T,  which equals almost exactly the respective critical fields obtained from the 2D fits of $C_{4f}(B)$. On closer inspection, an asymmetry of the zero-crossings, i. e., a larger positive and smaller negative wing for $x = 0.05$ and $0.1$, can be seen. This hints at the presence of a positive  additional  contribution to $\Gamma_{\rm mag}(B)$ peaked at $B = 2.5$~T. Indeed, the field dependence of  $\Gamma_{\rm mag}(B)$ for $x = 0.05$ above the critical field nicely follows the expected $1/(B - B_c)$ dependence for field-induced quantum criticality, cf. the blue line in
Fig.\ref{fig2}~(e). However, the low-field wing of the $\Gamma_{\rm mag}(B)$ curve deviates from $1/(B-B_c)$ (cf. black dotted line) due to an additional positive contribution to $\Gamma_{\rm mag}(B)$ between $2$ and $3$~T which is of similar size as the broad humps at larger $x$ (cf. Fig.~\ref{fig2}~(f)-(h)). This unexpected contribution, mostly located in the paramagnetic regime above $B_c$, therefore must have a different origin than quantum criticality \cite{Garst2005Sign}.
For $x=0.14$ no zero-crossing is found. However, the non-monotonic behavior of $\Gamma_{\rm mag}$ is compatible with a sign change around $B_c \approx 1$~T, concealed by a positive contribution with a maximum around $2.5$~T not captured by quantum criticality. This value of $B_c$ would be in agreement with the kink of $C_{4f}(B)/T$, cf. Fig.~\ref{fig2}~(c). Finally for $x = 0.16$ only a broad background, again with a maximum around $2.5$~T, is observed. This unexpected positive contribution to $\Gamma_{\rm mag} (B)$  is observed for all samples and peaks independent of $x$ at $B = 2.5$~T. In particular this contribution occurs independent of whether $B_c$ falls above or below. In fact, it is observed, mostly within the paramagnetic regime of quantum criticality and therefore must have a different origin than quantum criticality~\cite{Garst2005Sign}.

\begin{figure}[b]
\begin{center}
\includegraphics[width=0.85\linewidth]{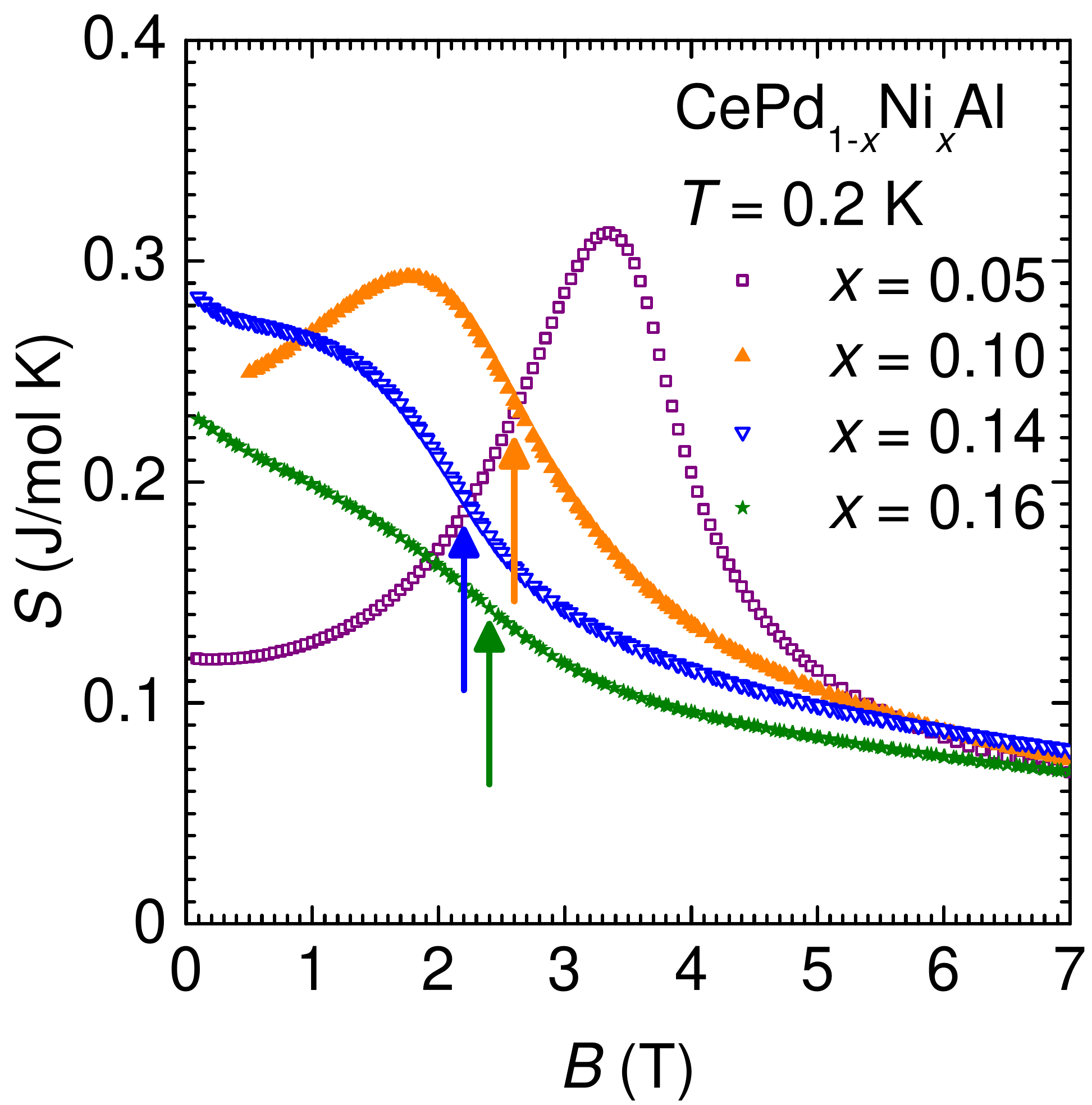}
\caption{$4f$ contribution to the entropy of CePd$_{1-x}$Ni$_x$Al (see text) vs. magnetic field $B$ at $T=0.2$~K for $B\|c$.   The colored arrows indicate the positions of anomalies in the magnetic Gr\"uneisen parameter (cf. Fig.~\ref{fig2}). }  \label{fig3}
\end{center}
\end{figure}

Using $\Gamma_{\rm mag}=-(dM/dT)/C$, the temperature derivative of the magnetization can be calculated from the data. The Maxwell relation
$dM/dT=dS/dB$ then allows to calculate the field dependence of the entropy by integration (SM) \cite{Supplement}.
Figure~\ref{fig3} shows the  entropy $S(B)$ for all four investigated single crystals of CePd$_{1-x}$Ni$_x$Al.   Entropy maxima are found for $x=0.05$ and $0.1$ near the respective critical fields. The colored arrows indicate the positions of the broad maxima in $\Gamma_{\rm mag}(B)$ (see respective arrows in Fig.~\ref{fig2}) and are close to inflection points of the respective entropy curves. A decrease of magnetic entropy generally points to a field-induced suppression of magnetic correlations.

\begin{figure}[t]
\begin{center}
\centerline{\includegraphics[width=\linewidth]{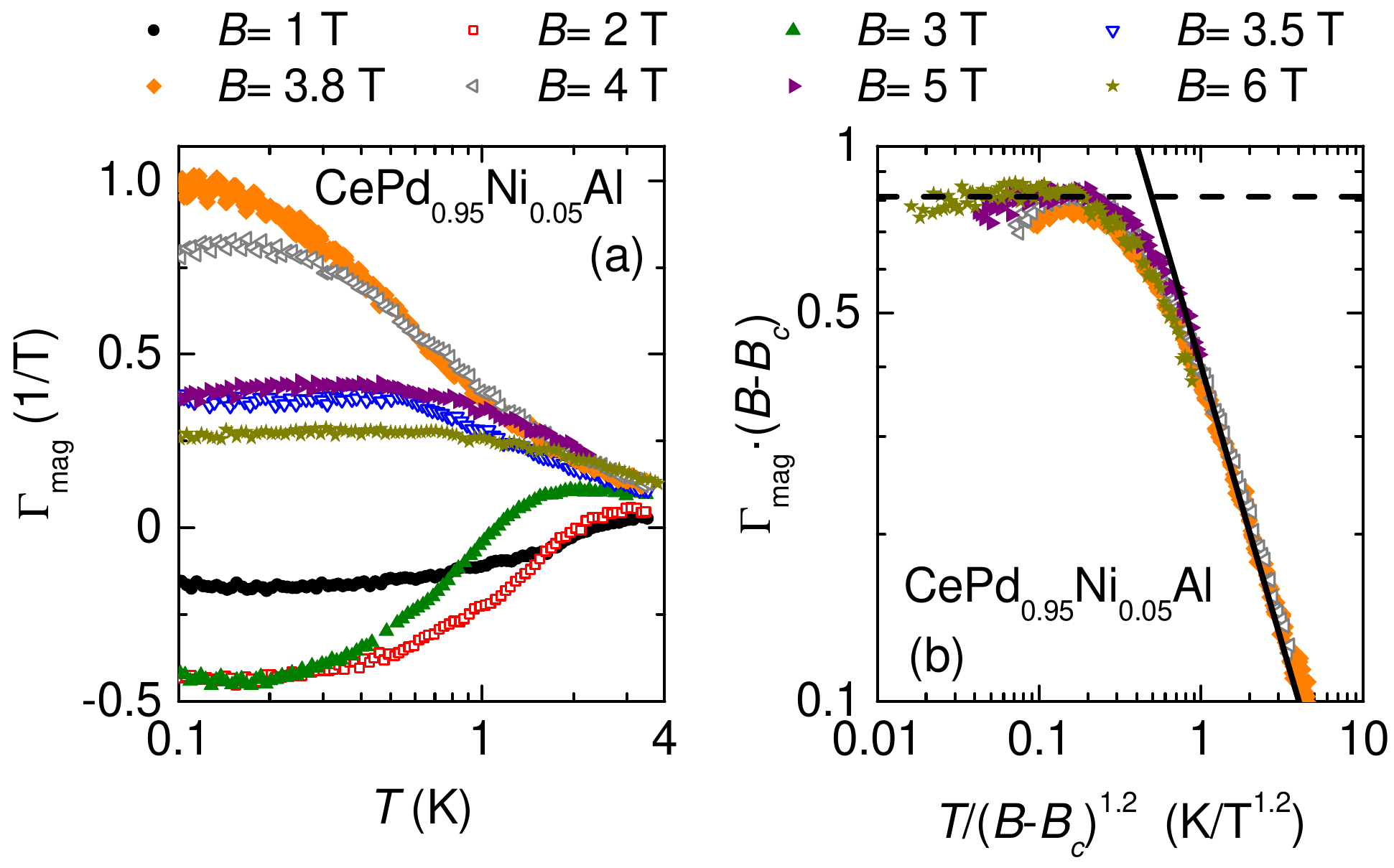}}
\caption{(a) $T$ dependence of $\Gamma_{\rm mag}$ for CePd$_{1-x}$Ni$_x$Al ($x=0.05$) at various fields along $c$ axis on semi-log scale.  (b)  Quantum critical scaling $\Gamma_{\rm mag} \cdot \left(B-B_c\right)$   vs. $T/(B-B_c)^{1.2}$ using $B_c=3$~T. The solid and dashed lines represent $\Gamma_{\rm mag} \propto \frac{1}{T}$ and $\Gamma_{\rm mag} =G_r$.}\label{fig4}
\end{center}
\end{figure}

Finally, we turn to an analysis of quantum criticality in $\Gamma_{\rm mag}$ for $x=0.05$. Scaling theory predicts a universal prefactor, $G_r = 0.5$, of the $\Gamma_{\rm mag}$ divergence for field-induced 3D AF quantum criticality \cite{Zhu2003Universally}. The value of $G_r = 0.81$ for $x = 0.05$ obtained from the fit in Fig.~\ref{fig2}~(e) deviates from this prediction. For 2D AF quantum criticality $G_r$ is non-universal and thus compatible with our data. Figure~\ref{fig4}~(a) shows the temperature dependence of $\Gamma_{\rm mag}$ at various fixed magnetic fields $B \parallel c$ while panel (b) shows a scaling plot of $\Gamma_{\rm mag} \cdot \left(B-B_c\right)$ vs $T/\left(B-B_c\right)^{1.2}$ for data at $B\geq 3.8$~T. The zero-crossings in panel (a) arise from the entropy accumulation at the phase boundary~\cite{Garst2005Sign}. The fact that the 3-T data display such a zero-crossing is consistent with the critical field of 3.4~T obtained from the heat capacity analysis. Quantum critical scaling holds for the data beyond the critical field (panel (b)).  Such scaling neglects the logarithmic correction $\log \frac{B_c}{B-B_c}$ predicted for 2D quantum criticality \cite{Zhu2003Universally}. Respectively, the best scaling is obtained for a somewhat low critical field of $3$~T.

Our thorough thermodynamic investigation of CePd$_{1-x}$Ni$_x$Al single crystals provides evidence for field-induced quantum criticality with the critical field being continuously suppressed towards zero for $x_c\approx 0.15$.  Previously, the equivalence of pressure and composition tuning has been demonstrated for CeCu$_{6-x}$Au$_x$~\cite{Loehneysen1996Non,Hamann2013Evolution} where unconventional quantum criticality with 2D critical fluctuations has been found~\cite{Stockert1998Two}. Suppressing, on the other hand, magnetic ordering for $x>x_c$  by a magnetic field, yields conventional quantum critical behavior corresponding to 3D fluctuations \cite{Loehneysen2001Pressure,Stockert2007Magnetic}. Our study of the heat capacity  of CePd$_{1-x}$Ni$_x$Al reveals 2D quantum criticality for field {\em and} concentration tuning at $x\geq 0.05$. This is thus the  second example for a material with 3D long-range order, displaying a reduced dimensionality of quantum critical fluctuations.  The strong geometrical frustration, leading to disordered moments separating chains of ordered moments, provides a natural explanation of 2D quantum criticality in CePd$_{1-x}$Ni$_x$Al~\cite{Doenni1996Geometrically,Nunez-Regueiro1997Magnetic}. Most interestingly, we have discovered an anomaly in the magnetic Gr\"uneisen parameter for  $x\geq 0.1$ that occurs in addition to the generic signatures of field-induced quantum criticality. For $x=0.05$ the deviation from generic $1/(B-B_c)$ behavior on the low-field side also hints at an additional contribution of similar order of magnitude in the same field range. Thus, this anomaly is independent of the AF QCP due to Ce(1) and Ce(3) moments and consequently ascribed to the frustrated Ce(2) moments. We note that the observed maximum of the magnetic Gr\"uneisen parameter is incompatible with the behavior of a heavy Fermi liquid composed of fully Kondo-screened Ce(2) moments. This indicates that magnetic frustration remains important as the material is concentration-tuned across the QCP even beyond $x_c$ as long as magnetic fluctuations lead to the non-equivalence of Ce moments.

We thank U. Burkhardt for the EDX measurements and M. Garst and K. Grube for fruitful discussions. This work was supported by the JSPS program for Postdoctoral Fellow Research Abroad, the German Science Foundation (FOR 960) and the Helmholtz Association (VI 521).

 \foreach \x in {1,2}
  {%
  \clearpage
    

\section*{Supplementary material}

Akito Sakai,[1] Stefan Lucas,[2] Philipp Gegenwart,[1] Oliver Stockert,[2] Hilbert v. Löhneysen,[3] and Veronika Fritsch[1]

[1]*Experimental Physics VI, Center for Electronic Correlations and Magnetism,*
*Institute of Physics, University of Augsburg, 86135 Augsburg, Germany*
[2]*Max Planck Institute for Chemical Physics of Solids, 01187 Dresden, Germany*
[3]*Karlsruhe Institute of Technology, Institute for Solid State Physics and Physics Institute, 76131 Karlsruhe, Germany*
(Dated: December 7, 2016)

## SUPPLEMENTAL MATERIAL

*Specific heat in magnetic fields.* Specific-heat data for CePd$_{1-x}$Ni$_x$Al single crystals in magnetic fields applied parallel to the $c$-axis are shown in Figs A and B. For $x = 0.14$ and $0.16$, $C_{4f}/T$ decreases monotonically with increasing field, indicating the stabilization of Fermi-liquid behavior. By contrast, for both $x = 0.05$ and $0.1$ $C_{4f}/T$ initially increases with field, then passes over a maximum with a subsequent decrease. This is characteristic for field-induced quantum criticality.

*Determination of the $4f$ contribution to the specific heat.* The $4f$ contribution to the specific heat $C_{4f}$ was determined by subtracting a phonon and conduction-electron contribution, as well a nuclear contribution from the measured specific heat $C$. In order to determine the former the specific heat of the non-magnetic reference compound LuPdAl was measured down to $T =$ 1.8 K and extrapolated to lower temperatures by assuming $C = \gamma T + \beta T^3$ with $\gamma = 5.6$ mJ/mol K$^2$ and $\beta = 0.72$ mJ/mol K$^4$.

The nuclear contribution was determined by assuming a field-independent Fermi-liquid (FL) state at high fields, i.e., $C/T = \gamma + A_n(B)/T^3$, and extrapolated to lower fields assuming $A_n(B)$ being linear in $B^2$. These $A_n$ values were then adjusted within the error bars of the fits, so that the subtraction did neither leave a nuclear-contribution-related upturn in $C_{4f}/T$ nor result in a physically meaningless peak in $C_{4f}/T$. The resulting parameters for $A_n$ are given in Tab. A. The quadrupolar contribution $A_n(B = 0)$ was determined independently without resorting to $A_n(B)$.

For the assignment of the error bars we assumed at low fields an experimental error of $\Delta C_{4f}/T = \pm 0.05$ J/mol K$^2$. At high fields the dominant contribution to the error comes from the nuclear contribution,

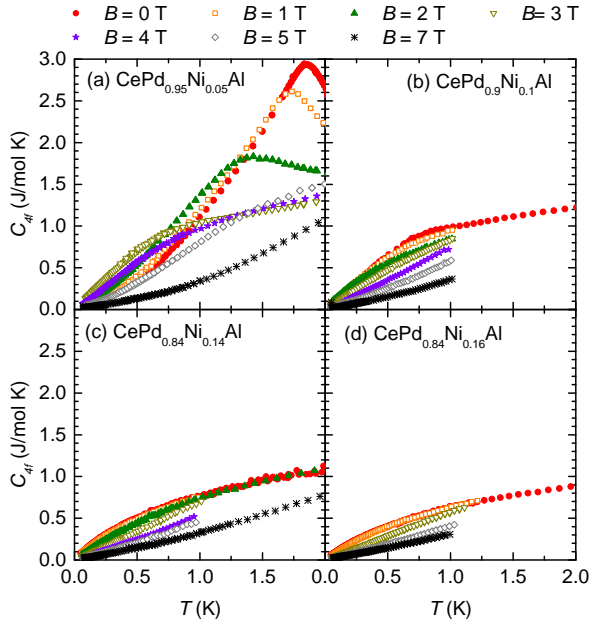

FIG. A. $4f$-electron contribution to the specific heat plotted as $C_{4f}$ vs. $T$ for various CePd$_{1-x}$Ni$_x$Al single crystals at $B \parallel c$ for (a) $x = 0.05$, (b) $x = 0.10$, (c) $x = 0.14$ and (d) $x = 0.16$.

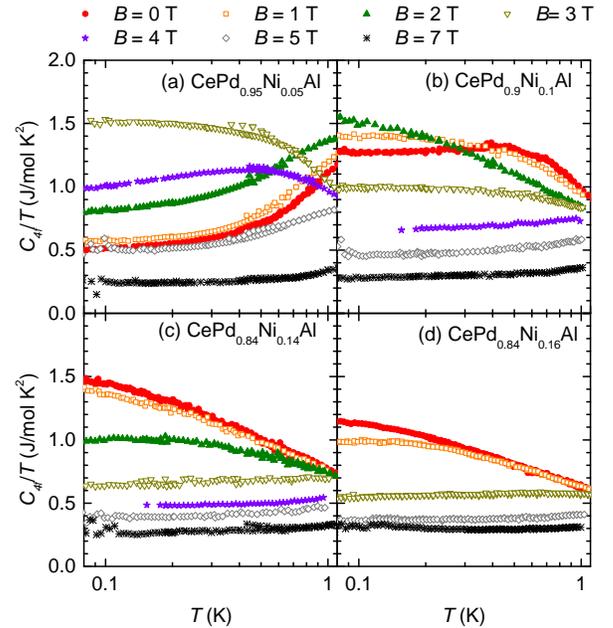

FIG. B. $4f$-electron contribution to the specific heat plotted as $C_{4f}/T$ vs. $T$ (on a logarithmic scale) for various CePd$_{1-x}$Ni$_x$Al single crystals at $B \parallel c$ for (a) $x = 0.05$, (b) $x = 0.10$, (c) $x = 0.14$ and (d) $x = 0.16$.

TABLE A. Parameters $A_n$ in units of J·K/mol used to determine the nuclear contribution to the specific heat.

|       | $x = 0.05$ | $x = 0.10$ | $x = 0.14$ | $x = 0.16$ |
|-------|---|---|---|---|
| $B = 0$   | $1.0 \cdot 10^{-5}$ | $1.3 \cdot 10^{-5}$ | $1.3 \cdot 10^{-5}$ | $2.0 \cdot 10^{-5}$ |
| $B = 1$ T | $2.0 \cdot 10^{-5}$ | $2.8 \cdot 10^{-5}$ | $2.6 \cdot 10^{-5}$ | $3.8 \cdot 10^{-5}$ |
| $B = 2$ T | $4.8 \cdot 10^{-5}$ | $4.0 \cdot 10^{-5}$ | $9.0 \cdot 10^{-5}$ |   |
| $B = 3$ T | $9.6 \cdot 10^{-5}$ | $9.4 \cdot 10^{-5}$ | $1.2 \cdot 10^{-4}$ | $9.2 \cdot 10^{-5}$ |
| $B = 4$ T | $1.6 \cdot 10^{-4}$ | $1.7 \cdot 10^{-4}$ | $2.0 \cdot 10^{-4}$ |   |
| $B = 5$ T | $2.5 \cdot 10^{-4}$ | $2.6 \cdot 10^{-4}$ | $2.9 \cdot 10^{-4}$ | $2.4 \cdot 10^{-4}$ |
| $B = 7$ T | $4.8 \cdot 10^{-4}$ | $5.0 \cdot 10^{-4}$ | $6.3 \cdot 10^{-4}$ | $4.6 \cdot 10^{-4}$ |

TABLE B. Ranges of $B_c$ compatible with the error bars of the $C/T$ data for the following functions: (a) 2D AF quantum criticality according to $C_{4f}/T = -a_1 \ln(a_2 b)$, $b = (B - B_c)/B_c$, (b) 3D AF quantum criticality according to $C_{4f}/T = \gamma_0 - c\sqrt{B - B_c}$.

|       |                 | $x = 0.05$ | $x = 0.10$ | $x = 0.14$ | $x = 0.16$ |
|-------|-----------------|---|---|---|---|
| (a)   | $B_c$ (T)       | 3.4 | 1.5 | 0.52 | $-1.55$ |
|       | lower limit (T) | 3.2 | 1.0 | 0.01 | $-0.5$  |
|       | upper limit (T) | 3.4 | 2.1 | 1.0  | $-5.1$  |
| (b)   | $B_c$ (T)       | 3.5 | 2.0 | 0.98 | $-0.23$ |
|       | lower limit (T) | 3.5 | 1.9 | 0.98 | $-0.8$  |
|       | upper limit (T) | 4   | 2.2 | 1.0  | $-0.05$ |

thus we assumed an error of 10 % of the nuclear contribution.

*Error bars of the fit parameters of Tab. I of the main paper* Due to the small number of datapoints for each fit in Fig. 2 (a)-(d) the statistical error for the fits is too large and thus meaningless. Instead the fits were repeated with fixed critical fields below and above the $B_c$ values found in Tab. I of the main paper and the range of $B_c$ compatible with the error bars of the $C/T$ data was determined. The results are given in Tab. B.

*Calculation of the field dependence of the magnetic entropy of $CePd_{1-x}Ni_xAl$ single crystals at 0.2 K.* The field-dependence of the specific heat was measured for $x = 0.05$ and obtained for the other $x$ by interpolating the data of $C$ at $T = 0.2$ K in the fields shown in Fig. B. From the field dependence of the specific heat $C$ and the magnetic Grüneisen parameter $\Gamma_{\mathrm{mag}}$ we can determine the field-dependence of the magnetic entropy. Since $\Gamma_{\mathrm{mag}} = -\frac{dM/dT}{C}$ we can calculate $\frac{dM}{dT}\big|_{B=\mathrm{const.}} = \frac{dS}{dB}\big|_{T=\mathrm{const.}}$ via the Ehrenfest relation.

By integration, we obtain the field-dependent entropy

$$S = S_0 + \Delta S = S_0 + \int_0^B \frac{dS}{dB'} dB',$$

where $S_0$ is the zero-field entropy at $T = 200$ mK.

As shown in the main paper the specific heat $C$ of all

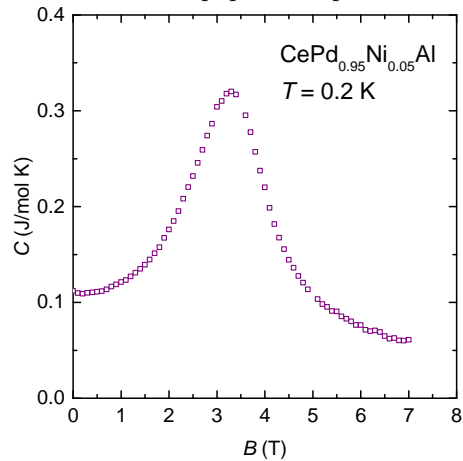

FIG. C. Specific heat $C$ of $CePd_{0.95}Ni_{0.05}Al$ at $T = 0.2$ K vs. magnetic field $B\|c$.

samples was measured down to $T \approx 70$ mK. We extrapolated the data down to 10 mK. From the resulting curve $C/T$ was calculated and integrated from 10 mK to obtain the zero-field entropy.

Figure C displays the specific heat $C$ vs. magnetic field $B$ at $T = 0.2$ K of $CePd_{0.95}Ni_{0.05}Al$. Subsequently the specific heat data were multiplied with $-\Gamma_{\mathrm{mag}}$ obtained from Figs 3 (e)-(h) of the main paper. The product equals $\frac{dS}{dB}$. Integration over the magnetic field resulted in $\Delta S = \int_{B_0}^{B} \frac{dS}{dB'} dB'$, with $B_0$ being the lowest field where data of $\Gamma_{\mathrm{mag}}$ were available, and hence the entropy shown in Fig. 4 of the main paper.


  }


\begin{thebibliography}{45}%
\makeatletter
\providecommand \@ifxundefined [1]{%
 \@ifx{#1\undefined}
}%
\providecommand \@ifnum [1]{%
 \ifnum #1\expandafter \@firstoftwo
 \else \expandafter \@secondoftwo
 \fi
}%
\providecommand \@ifx [1]{%
 \ifx #1\expandafter \@firstoftwo
 \else \expandafter \@secondoftwo
 \fi
}%
\providecommand \natexlab [1]{#1}%
\providecommand \enquote  [1]{``#1''}%
\providecommand \bibnamefont  [1]{#1}%
\providecommand \bibfnamefont [1]{#1}%
\providecommand \citenamefont [1]{#1}%
\providecommand \href@noop [0]{\@secondoftwo}%
\providecommand \href [0]{\begingroup \@sanitize@url \@href}%
\providecommand \@href[1]{\@@startlink{#1}\@@href}%
\providecommand \@@href[1]{\endgroup#1\@@endlink}%
\providecommand \@sanitize@url [0]{\catcode `\\12\catcode `\$12\catcode
  `\&12\catcode `\#12\catcode `\^12\catcode `\_12\catcode `\%12\relax}%
\providecommand \@@startlink[1]{}%
\providecommand \@@endlink[0]{}%
\providecommand \url  [0]{\begingroup\@sanitize@url \@url }%
\providecommand \@url [1]{\endgroup\@href {#1}{\urlprefix }}%
\providecommand \urlprefix  [0]{URL }%
\providecommand \Eprint [0]{\href }%
\providecommand \doibase [0]{http://dx.doi.org/}%
\providecommand \selectlanguage [0]{\@gobble}%
\providecommand \bibinfo  [0]{\@secondoftwo}%
\providecommand \bibfield  [0]{\@secondoftwo}%
\providecommand \translation [1]{[#1]}%
\providecommand \BibitemOpen [0]{}%
\providecommand \bibitemStop [0]{}%
\providecommand \bibitemNoStop [0]{.\EOS\space}%
\providecommand \EOS [0]{\spacefactor3000\relax}%
\providecommand \BibitemShut  [1]{\csname bibitem#1\endcsname}%
\let\auto@bib@innerbib\@empty
\bibitem [{\citenamefont {Mathur}\ \emph {et~al.}(1998)\citenamefont {Mathur},
  \citenamefont {Grosche}, \citenamefont {Julian}, \citenamefont {Walker},
  \citenamefont {Freye}, \citenamefont {Haselwimmer},\ and\ \citenamefont
  {Lonzarich}}]{Mathur1998Magnetically}%
  \BibitemOpen
  \bibfield  {author} {\bibinfo {author} {\bibfnamefont {N.}~\bibnamefont
  {Mathur}}, \bibinfo {author} {\bibfnamefont {F.}~\bibnamefont {Grosche}},
  \bibinfo {author} {\bibfnamefont {S.}~\bibnamefont {Julian}}, \bibinfo
  {author} {\bibfnamefont {I.}~\bibnamefont {Walker}}, \bibinfo {author}
  {\bibfnamefont {D.}~\bibnamefont {Freye}}, \bibinfo {author} {\bibfnamefont
  {R.}~\bibnamefont {Haselwimmer}}, \ and\ \bibinfo {author} {\bibfnamefont
  {G.}~\bibnamefont {Lonzarich}},\ }\href@noop {} {\bibfield  {journal}
  {\bibinfo  {journal} {Nature}\ }\textbf {\bibinfo {volume} {394}},\ \bibinfo
  {pages} {39} (\bibinfo {year} {1998})}\BibitemShut {NoStop}%
\bibitem [{\citenamefont {Ramshaw}\ \emph {et~al.}(2015)\citenamefont
  {Ramshaw}, \citenamefont {Sebastian}, \citenamefont {McDonald}, \citenamefont
  {Day}, \citenamefont {Tan}, \citenamefont {Zhu}, \citenamefont {Betts},
  \citenamefont {Liang}, \citenamefont {Bonn}, \citenamefont {Hardy} \emph
  {et~al.}}]{Ramshaw2015Quasiparticle}%
  \BibitemOpen
  \bibfield  {author} {\bibinfo {author} {\bibfnamefont {B.}~\bibnamefont
  {Ramshaw}}, \bibinfo {author} {\bibfnamefont {S.}~\bibnamefont {Sebastian}},
  \bibinfo {author} {\bibfnamefont {R.}~\bibnamefont {McDonald}}, \bibinfo
  {author} {\bibfnamefont {J.}~\bibnamefont {Day}}, \bibinfo {author}
  {\bibfnamefont {B.}~\bibnamefont {Tan}}, \bibinfo {author} {\bibfnamefont
  {Z.}~\bibnamefont {Zhu}}, \bibinfo {author} {\bibfnamefont {J.}~\bibnamefont
  {Betts}}, \bibinfo {author} {\bibfnamefont {R.}~\bibnamefont {Liang}},
  \bibinfo {author} {\bibfnamefont {D.}~\bibnamefont {Bonn}}, \bibinfo {author}
  {\bibfnamefont {W.}~\bibnamefont {Hardy}},  \emph {et~al.},\ }\href@noop {}
  {\bibfield  {journal} {\bibinfo  {journal} {Science}\ }\textbf {\bibinfo
  {volume} {348}},\ \bibinfo {pages} {317} (\bibinfo {year}
  {2015})}\BibitemShut {NoStop}%
\bibitem [{\citenamefont {Steglich}\ \emph {et~al.}(1979)\citenamefont
  {Steglich}, \citenamefont {Aarts}, \citenamefont {Bredl}, \citenamefont
  {Lieke}, \citenamefont {Meschede}, \citenamefont {Franz},\ and\ \citenamefont
  {Sch\"{a}fer}}]{Steglich1979Superconductivity}%
  \BibitemOpen
  \bibfield  {author} {\bibinfo {author} {\bibfnamefont {F.}~\bibnamefont
  {Steglich}}, \bibinfo {author} {\bibfnamefont {J.}~\bibnamefont {Aarts}},
  \bibinfo {author} {\bibfnamefont {C.}~\bibnamefont {Bredl}}, \bibinfo
  {author} {\bibfnamefont {W.}~\bibnamefont {Lieke}}, \bibinfo {author}
  {\bibfnamefont {D.}~\bibnamefont {Meschede}}, \bibinfo {author}
  {\bibfnamefont {W.}~\bibnamefont {Franz}}, \ and\ \bibinfo {author}
  {\bibfnamefont {H.}~\bibnamefont {Sch\"{a}fer}},\ }\href@noop {} {\bibfield
  {journal} {\bibinfo  {journal} {Phys. Rev. Lett.}\ }\textbf {\bibinfo
  {volume} {43}},\ \bibinfo {pages} {1892} (\bibinfo {year}
  {1979})}\BibitemShut {NoStop}%
\bibitem [{\citenamefont {Sachdev}(2010)}]{Sachdev2010Where}%
  \BibitemOpen
  \bibfield  {author} {\bibinfo {author} {\bibfnamefont {S.}~\bibnamefont
  {Sachdev}},\ }\href {\doibase 10.1002/pssb.200983037} {\bibfield  {journal}
  {\bibinfo  {journal} {Phys. Status Solidi B}\ }\textbf {\bibinfo {volume}
  {247}},\ \bibinfo {pages} {537} (\bibinfo {year} {2010})}\BibitemShut
  {NoStop}%
\bibitem [{\citenamefont {Stewart}(2011)}]{Stewart2011Superconductivity}%
  \BibitemOpen
  \bibfield  {author} {\bibinfo {author} {\bibfnamefont {G.~R.}\ \bibnamefont
  {Stewart}},\ }\href {\doibase 10.1103/RevModPhys.83.1589} {\bibfield
  {journal} {\bibinfo  {journal} {Rev. Mod. Phys.}\ }\textbf {\bibinfo {volume}
  {83}},\ \bibinfo {pages} {1589} (\bibinfo {year} {2011})}\BibitemShut
  {NoStop}%
\bibitem [{\citenamefont {Ramirez}(1994)}]{Ramirez1994Strongly}%
  \BibitemOpen
  \bibfield  {author} {\bibinfo {author} {\bibfnamefont {A.~P.}\ \bibnamefont
  {Ramirez}},\ }\href@noop {} {\bibfield  {journal} {\bibinfo  {journal} {Annu.
  Rev. Mater. Sci.}\ }\textbf {\bibinfo {volume} {24}},\ \bibinfo {pages} {453}
  (\bibinfo {year} {1994})}\BibitemShut {NoStop}%
\bibitem [{\citenamefont {Balents}(2010)}]{Balents2010Spin}%
  \BibitemOpen
  \bibfield  {author} {\bibinfo {author} {\bibfnamefont {L.}~\bibnamefont
  {Balents}},\ }\href@noop {} {\bibfield  {journal} {\bibinfo  {journal}
  {Nature}\ }\textbf {\bibinfo {volume} {464}},\ \bibinfo {pages} {199}
  (\bibinfo {year} {2010})}\BibitemShut {NoStop}%
\bibitem [{\citenamefont {Fritsch}\ \emph {et~al.}(2006)\citenamefont
  {Fritsch}, \citenamefont {Thompson}, \citenamefont {Sarrao}, \citenamefont
  {{Krug von Nidda}}, \citenamefont {Eremina},\ and\ \citenamefont
  {Loidl}}]{Fritsch2006Correlation}%
  \BibitemOpen
  \bibfield  {author} {\bibinfo {author} {\bibfnamefont {V.}~\bibnamefont
  {Fritsch}}, \bibinfo {author} {\bibfnamefont {J.~D.}\ \bibnamefont
  {Thompson}}, \bibinfo {author} {\bibfnamefont {J.~L.}\ \bibnamefont
  {Sarrao}}, \bibinfo {author} {\bibfnamefont {H.-A.}\ \bibnamefont {{Krug von
  Nidda}}}, \bibinfo {author} {\bibfnamefont {R.~M.}\ \bibnamefont {Eremina}},
  \ and\ \bibinfo {author} {\bibfnamefont {A.}~\bibnamefont {Loidl}},\
  }\href@noop {} {\bibfield  {journal} {\bibinfo  {journal} {Phys. Rev. B}\
  }\textbf {\bibinfo {volume} {73}},\ \bibinfo {pages} {094413} (\bibinfo
  {year} {2006})}\BibitemShut {NoStop}%
\bibitem [{\citenamefont {Doniach}(1977)}]{Doniach1977Kondo}%
  \BibitemOpen
  \bibfield  {author} {\bibinfo {author} {\bibfnamefont {S.}~\bibnamefont
  {Doniach}},\ }\href@noop {} {\bibfield  {journal} {\bibinfo  {journal}
  {Physica B}\ }\textbf {\bibinfo {volume} {91}},\ \bibinfo {pages} {231 }
  (\bibinfo {year} {1977})}\BibitemShut {NoStop}%
\bibitem [{\citenamefont {Stewart}(2001)}]{Stewart2001Non}%
  \BibitemOpen
  \bibfield  {author} {\bibinfo {author} {\bibfnamefont {G.~R.}\ \bibnamefont
  {Stewart}},\ }\href@noop {} {\bibfield  {journal} {\bibinfo  {journal} {Rev.
  Mod. Phys.}\ }\textbf {\bibinfo {volume} {73}},\ \bibinfo {pages} {797}
  (\bibinfo {year} {2001})}\BibitemShut {NoStop}%
\bibitem [{\citenamefont {v.~L\"{o}hneysen}\ \emph {et~al.}(2007)\citenamefont
  {v.~L\"{o}hneysen}, \citenamefont {Rosch}, \citenamefont {Vojta},\ and\
  \citenamefont {W\"{o}lfle}}]{Loehneysen2007Fermi}%
  \BibitemOpen
  \bibfield  {author} {\bibinfo {author} {\bibfnamefont {H.}~\bibnamefont
  {v.~L\"{o}hneysen}}, \bibinfo {author} {\bibfnamefont {A.}~\bibnamefont
  {Rosch}}, \bibinfo {author} {\bibfnamefont {M.}~\bibnamefont {Vojta}}, \ and\
  \bibinfo {author} {\bibfnamefont {P.}~\bibnamefont {W\"{o}lfle}},\
  }\href@noop {} {\bibfield  {journal} {\bibinfo  {journal} {Rev. Mod. Phys.}\
  }\textbf {\bibinfo {volume} {79}},\ \bibinfo {pages} {1015} (\bibinfo {year}
  {2007})}\BibitemShut {NoStop}%
\bibitem [{\citenamefont {Gegenwart}\ \emph {et~al.}(2008)\citenamefont
  {Gegenwart}, \citenamefont {Si},\ and\ \citenamefont
  {Steglich}}]{Gegenwart2008Quantum}%
  \BibitemOpen
  \bibfield  {author} {\bibinfo {author} {\bibfnamefont {P.}~\bibnamefont
  {Gegenwart}}, \bibinfo {author} {\bibfnamefont {Q.}~\bibnamefont {Si}}, \
  and\ \bibinfo {author} {\bibfnamefont {F.}~\bibnamefont {Steglich}},\
  }\href@noop {} {\bibfield  {journal} {\bibinfo  {journal} {Nat. Phys.}\
  }\textbf {\bibinfo {volume} {4}},\ \bibinfo {pages} {186} (\bibinfo {year}
  {2008})}\BibitemShut {NoStop}%
\bibitem [{\citenamefont {Senthil}\ \emph {et~al.}(2004)\citenamefont
  {Senthil}, \citenamefont {Vojta},\ and\ \citenamefont
  {Sachdev}}]{Senthil2004Weak}%
  \BibitemOpen
  \bibfield  {author} {\bibinfo {author} {\bibfnamefont {T.}~\bibnamefont
  {Senthil}}, \bibinfo {author} {\bibfnamefont {M.}~\bibnamefont {Vojta}}, \
  and\ \bibinfo {author} {\bibfnamefont {S.}~\bibnamefont {Sachdev}},\ }\href
  {\doibase 10.1103/PhysRevB.69.035111} {\bibfield  {journal} {\bibinfo
  {journal} {Phys. Rev. B}\ }\textbf {\bibinfo {volume} {69}},\ \bibinfo
  {pages} {035111} (\bibinfo {year} {2004})}\BibitemShut {NoStop}%
\bibitem [{\citenamefont {Si}(2006)}]{Si2006Global}%
  \BibitemOpen
  \bibfield  {author} {\bibinfo {author} {\bibfnamefont {Q.}~\bibnamefont
  {Si}},\ }\href@noop {} {\bibfield  {journal} {\bibinfo  {journal} {Physica
  B}\ }\textbf {\bibinfo {volume} {378-380}},\ \bibinfo {pages} {23} (\bibinfo
  {year} {2006})}\BibitemShut {NoStop}%
\bibitem [{\citenamefont {Vojta}(2008)}]{Vojta2008From}%
  \BibitemOpen
  \bibfield  {author} {\bibinfo {author} {\bibfnamefont {M.}~\bibnamefont
  {Vojta}},\ }\href@noop {} {\bibfield  {journal} {\bibinfo  {journal} {Phys.
  Rev. B}\ }\textbf {\bibinfo {volume} {78}},\ \bibinfo {pages} {125109}
  (\bibinfo {year} {2008})}\BibitemShut {NoStop}%
\bibitem [{\citenamefont {Coleman}\ and\ \citenamefont
  {Nevidomskyy}(2010)}]{Coleman2010Frustration}%
  \BibitemOpen
  \bibfield  {author} {\bibinfo {author} {\bibfnamefont {P.}~\bibnamefont
  {Coleman}}\ and\ \bibinfo {author} {\bibfnamefont {A.}~\bibnamefont
  {Nevidomskyy}},\ }\href@noop {} {\bibfield  {journal} {\bibinfo  {journal}
  {J. Low Temp. Phys.}\ }\textbf {\bibinfo {volume} {161}},\ \bibinfo {pages}
  {182} (\bibinfo {year} {2010})}\BibitemShut {NoStop}%
\bibitem [{\citenamefont {Bud'ko}\ \emph {et~al.}(2004)\citenamefont {Bud'ko},
  \citenamefont {Morosan},\ and\ \citenamefont {Canfield}}]{Budko2004Magnetic}%
  \BibitemOpen
  \bibfield  {author} {\bibinfo {author} {\bibfnamefont {S.~L.}\ \bibnamefont
  {Bud'ko}}, \bibinfo {author} {\bibfnamefont {E.}~\bibnamefont {Morosan}}, \
  and\ \bibinfo {author} {\bibfnamefont {P.~C.}\ \bibnamefont {Canfield}},\
  }\href@noop {} {\bibfield  {journal} {\bibinfo  {journal} {Phys. Rev. B}\
  }\textbf {\bibinfo {volume} {69}},\ \bibinfo {pages} {014415} (\bibinfo
  {year} {2004})}\BibitemShut {NoStop}%
\bibitem [{\citenamefont {Tokiwa}\ \emph {et~al.}(2013)\citenamefont {Tokiwa},
  \citenamefont {Garst}, \citenamefont {Gegenwart}, \citenamefont {Bud'ko},\
  and\ \citenamefont {Canfield}}]{Tokiwa2013Quantum}%
  \BibitemOpen
  \bibfield  {author} {\bibinfo {author} {\bibfnamefont {Y.}~\bibnamefont
  {Tokiwa}}, \bibinfo {author} {\bibfnamefont {M.}~\bibnamefont {Garst}},
  \bibinfo {author} {\bibfnamefont {P.}~\bibnamefont {Gegenwart}}, \bibinfo
  {author} {\bibfnamefont {S.~L.}\ \bibnamefont {Bud'ko}}, \ and\ \bibinfo
  {author} {\bibfnamefont {P.~C.}\ \bibnamefont {Canfield}},\ }\href@noop {}
  {\bibfield  {journal} {\bibinfo  {journal} {Phys. Rev. Lett.}\ }\textbf
  {\bibinfo {volume} {111}},\ \bibinfo {pages} {116401} (\bibinfo {year}
  {2013})}\BibitemShut {NoStop}%
\bibitem [{\citenamefont {Dong}\ \emph {et~al.}(2013)\citenamefont {Dong},
  \citenamefont {Tokiwa}, \citenamefont {Bud'ko}, \citenamefont {Canfield},\
  and\ \citenamefont {Gegenwart}}]{Dong2013Anomalous}%
  \BibitemOpen
  \bibfield  {author} {\bibinfo {author} {\bibfnamefont {J.~K.}\ \bibnamefont
  {Dong}}, \bibinfo {author} {\bibfnamefont {Y.}~\bibnamefont {Tokiwa}},
  \bibinfo {author} {\bibfnamefont {S.~L.}\ \bibnamefont {Bud'ko}}, \bibinfo
  {author} {\bibfnamefont {P.~C.}\ \bibnamefont {Canfield}}, \ and\ \bibinfo
  {author} {\bibfnamefont {P.}~\bibnamefont {Gegenwart}},\ }\href {\doibase
  10.1103/PhysRevLett.110.176402} {\bibfield  {journal} {\bibinfo  {journal}
  {Phys. Rev. Lett.}\ }\textbf {\bibinfo {volume} {110}},\ \bibinfo {pages}
  {176402} (\bibinfo {year} {2013})}\BibitemShut {NoStop}%
\bibitem [{\citenamefont {Tokiwa}\ \emph {et~al.}(2015)\citenamefont {Tokiwa},
  \citenamefont {Stingl}, \citenamefont {Kim}, \citenamefont {Takabatake},\
  and\ \citenamefont {Gegenwart}}]{Tokiwa2015Characteristic}%
  \BibitemOpen
  \bibfield  {author} {\bibinfo {author} {\bibfnamefont {Y.}~\bibnamefont
  {Tokiwa}}, \bibinfo {author} {\bibfnamefont {C.}~\bibnamefont {Stingl}},
  \bibinfo {author} {\bibfnamefont {M.-S.}\ \bibnamefont {Kim}}, \bibinfo
  {author} {\bibfnamefont {T.}~\bibnamefont {Takabatake}}, \ and\ \bibinfo
  {author} {\bibfnamefont {P.}~\bibnamefont {Gegenwart}},\ }\href@noop {}
  {\bibfield  {journal} {\bibinfo  {journal} {Sci. Adv.}\ }\textbf {\bibinfo
  {volume} {1}},\ \bibinfo {pages} {e1500001} (\bibinfo {year}
  {2015})}\BibitemShut {NoStop}%
\bibitem [{\citenamefont {{D\"{o}nni}}\ \emph {et~al.}(1996)\citenamefont
  {{D\"{o}nni}}, \citenamefont {Ehlers}, \citenamefont {Maletta}, \citenamefont
  {Fischer}, \citenamefont {Kitazawa},\ and\ \citenamefont
  {Zolliker}}]{Doenni1996Geometrically}%
  \BibitemOpen
  \bibfield  {author} {\bibinfo {author} {\bibfnamefont {A.}~\bibnamefont
  {{D\"{o}nni}}}, \bibinfo {author} {\bibfnamefont {G.}~\bibnamefont {Ehlers}},
  \bibinfo {author} {\bibfnamefont {H.}~\bibnamefont {Maletta}}, \bibinfo
  {author} {\bibfnamefont {P.}~\bibnamefont {Fischer}}, \bibinfo {author}
  {\bibfnamefont {H.}~\bibnamefont {Kitazawa}}, \ and\ \bibinfo {author}
  {\bibfnamefont {M.}~\bibnamefont {Zolliker}},\ }\href@noop {} {\bibfield
  {journal} {\bibinfo  {journal} {J. Phys.: Cond. Mat.}\ }\textbf {\bibinfo
  {volume} {8}},\ \bibinfo {pages} {11213} (\bibinfo {year}
  {1996})}\BibitemShut {NoStop}%
\bibitem [{\citenamefont {Oyamada}\ \emph {et~al.}(1996)\citenamefont
  {Oyamada}, \citenamefont {Kamioka}, \citenamefont {Hashi}, \citenamefont
  {Maegawa}, \citenamefont {Goto},\ and\ \citenamefont
  {Kitazawa}}]{Oyamada1996NMR}%
  \BibitemOpen
  \bibfield  {author} {\bibinfo {author} {\bibfnamefont {A.}~\bibnamefont
  {Oyamada}}, \bibinfo {author} {\bibfnamefont {K.}~\bibnamefont {Kamioka}},
  \bibinfo {author} {\bibfnamefont {K.}~\bibnamefont {Hashi}}, \bibinfo
  {author} {\bibfnamefont {S.}~\bibnamefont {Maegawa}}, \bibinfo {author}
  {\bibfnamefont {T.}~\bibnamefont {Goto}}, \ and\ \bibinfo {author}
  {\bibfnamefont {H.}~\bibnamefont {Kitazawa}},\ }\href@noop {} {\bibfield
  {journal} {\bibinfo  {journal} {J. Phys. Soc. Jpn.}\ }\textbf {\bibinfo
  {volume} {65}},\ \bibinfo {pages} {128} (\bibinfo {year} {1996})}\BibitemShut
  {NoStop}%
\bibitem [{\citenamefont {Lacroix}(2010)}]{Lacroix2010Frustrated}%
  \BibitemOpen
  \bibfield  {author} {\bibinfo {author} {\bibfnamefont {C.}~\bibnamefont
  {Lacroix}},\ }\href@noop {} {\bibfield  {journal} {\bibinfo  {journal} {J.
  Phys. Soc. Jpn.}\ }\textbf {\bibinfo {volume} {79}},\ \bibinfo {pages}
  {011008} (\bibinfo {year} {2010})}\BibitemShut {NoStop}%
\bibitem [{\citenamefont {Motome}\ \emph {et~al.}(2010)\citenamefont {Motome},
  \citenamefont {Nakamikawa}, \citenamefont {Yamaji},\ and\ \citenamefont
  {Udagawa}}]{Motome2010Partial}%
  \BibitemOpen
  \bibfield  {author} {\bibinfo {author} {\bibfnamefont {Y.}~\bibnamefont
  {Motome}}, \bibinfo {author} {\bibfnamefont {K.}~\bibnamefont {Nakamikawa}},
  \bibinfo {author} {\bibfnamefont {Y.}~\bibnamefont {Yamaji}}, \ and\ \bibinfo
  {author} {\bibfnamefont {M.}~\bibnamefont {Udagawa}},\ }\href@noop {}
  {\bibfield  {journal} {\bibinfo  {journal} {Phys. Rev. Lett.}\ }\textbf
  {\bibinfo {volume} {105}},\ \bibinfo {pages} {036403} (\bibinfo {year}
  {2010})}\BibitemShut {NoStop}%
\bibitem [{\citenamefont {Fritsch}\ \emph {et~al.}(2014)\citenamefont
  {Fritsch}, \citenamefont {Bagrets}, \citenamefont {Goll}, \citenamefont
  {Kittler}, \citenamefont {Wolf}, \citenamefont {Grube}, \citenamefont
  {Huang},\ and\ \citenamefont {{L\"ohneysen}}}]{Fritsch2014Approaching}%
  \BibitemOpen
  \bibfield  {author} {\bibinfo {author} {\bibfnamefont {V.}~\bibnamefont
  {Fritsch}}, \bibinfo {author} {\bibfnamefont {N.}~\bibnamefont {Bagrets}},
  \bibinfo {author} {\bibfnamefont {G.}~\bibnamefont {Goll}}, \bibinfo {author}
  {\bibfnamefont {W.}~\bibnamefont {Kittler}}, \bibinfo {author} {\bibfnamefont
  {M.~J.}\ \bibnamefont {Wolf}}, \bibinfo {author} {\bibfnamefont
  {K.}~\bibnamefont {Grube}}, \bibinfo {author} {\bibfnamefont {C.-L.}\
  \bibnamefont {Huang}}, \ and\ \bibinfo {author} {\bibfnamefont {H.~v.}\
  \bibnamefont {{L\"ohneysen}}},\ }\href {\doibase 10.1103/PhysRevB.89.054416}
  {\bibfield  {journal} {\bibinfo  {journal} {Phys. Rev. B}\ }\textbf {\bibinfo
  {volume} {89}},\ \bibinfo {pages} {054416} (\bibinfo {year}
  {2014})}\BibitemShut {NoStop}%
\bibitem [{\citenamefont {Isikawa}\ \emph {et~al.}(1996)\citenamefont
  {Isikawa}, \citenamefont {Mizushima}, \citenamefont {Fukushima},
  \citenamefont {Kuwai}, \citenamefont {Sakurai},\ and\ \citenamefont
  {Kitzawa}}]{Isikawa1996Magnetocrystalline}%
  \BibitemOpen
  \bibfield  {author} {\bibinfo {author} {\bibfnamefont {Y.}~\bibnamefont
  {Isikawa}}, \bibinfo {author} {\bibfnamefont {T.}~\bibnamefont {Mizushima}},
  \bibinfo {author} {\bibfnamefont {N.}~\bibnamefont {Fukushima}}, \bibinfo
  {author} {\bibfnamefont {T.}~\bibnamefont {Kuwai}}, \bibinfo {author}
  {\bibfnamefont {J.}~\bibnamefont {Sakurai}}, \ and\ \bibinfo {author}
  {\bibfnamefont {H.}~\bibnamefont {Kitzawa}},\ }\href@noop {} {\bibfield
  {journal} {\bibinfo  {journal} {J. Phys. Soc. Jpn.}\ }\textbf {\bibinfo
  {volume} {65 {\normalfont Suppl. B}}},\ \bibinfo {pages} {117} (\bibinfo
  {year} {1996})}\BibitemShut {NoStop}%
\bibitem [{\citenamefont {Kitazawa}\ \emph {et~al.}(1994)\citenamefont
  {Kitazawa}, \citenamefont {Matsushita}, \citenamefont {Matsumoto},\ and\
  \citenamefont {Suzuki}}]{Kitazawa1994Electronic}%
  \BibitemOpen
  \bibfield  {author} {\bibinfo {author} {\bibfnamefont {H.}~\bibnamefont
  {Kitazawa}}, \bibinfo {author} {\bibfnamefont {A.}~\bibnamefont
  {Matsushita}}, \bibinfo {author} {\bibfnamefont {T.}~\bibnamefont
  {Matsumoto}}, \ and\ \bibinfo {author} {\bibfnamefont {T.}~\bibnamefont
  {Suzuki}},\ }\href {\doibase http://dx.doi.org/10.1016/0921-4526(94)91726-4}
  {\bibfield  {journal} {\bibinfo  {journal} {Physica B}\ }\textbf {\bibinfo
  {volume} {199}},\ \bibinfo {pages} {28 } (\bibinfo {year}
  {1994})}\BibitemShut {NoStop}%
\bibitem [{\citenamefont {Fritsch}\ \emph {et~al.}(2015)\citenamefont
  {Fritsch}, \citenamefont {Stockert}, \citenamefont {Huang}, \citenamefont
  {Bagrets}, \citenamefont {Kittler}, \citenamefont {Taubenheim}, \citenamefont
  {Pilawa}, \citenamefont {Woitschach}, \citenamefont {Huesges}, \citenamefont
  {Lucas}, \citenamefont {Schneidewind}, \citenamefont {Grube},\ and\
  \citenamefont {{L\"ohneysen}}}]{Fritsch2015Role}%
  \BibitemOpen
  \bibfield  {author} {\bibinfo {author} {\bibfnamefont {V.}~\bibnamefont
  {Fritsch}}, \bibinfo {author} {\bibfnamefont {O.}~\bibnamefont {Stockert}},
  \bibinfo {author} {\bibfnamefont {C.-L.}\ \bibnamefont {Huang}}, \bibinfo
  {author} {\bibfnamefont {N.}~\bibnamefont {Bagrets}}, \bibinfo {author}
  {\bibfnamefont {W.}~\bibnamefont {Kittler}}, \bibinfo {author} {\bibfnamefont
  {C.}~\bibnamefont {Taubenheim}}, \bibinfo {author} {\bibfnamefont
  {B.}~\bibnamefont {Pilawa}}, \bibinfo {author} {\bibfnamefont
  {S.}~\bibnamefont {Woitschach}}, \bibinfo {author} {\bibfnamefont
  {Z.}~\bibnamefont {Huesges}}, \bibinfo {author} {\bibfnamefont
  {S.}~\bibnamefont {Lucas}}, \bibinfo {author} {\bibfnamefont
  {A.}~\bibnamefont {Schneidewind}}, \bibinfo {author} {\bibfnamefont
  {K.}~\bibnamefont {Grube}}, \ and\ \bibinfo {author} {\bibfnamefont {H.~v.}\
  \bibnamefont {{L\"ohneysen}}},\ }\href@noop {} {\bibfield  {journal}
  {\bibinfo  {journal} {Eur. Phys. J. ST}\ }\textbf {\bibinfo {volume} {224}},\
  \bibinfo {pages} {997–1019} (\bibinfo {year} {2015})}\BibitemShut {NoStop}%
\bibitem [{\citenamefont {Huo}\ \emph {et~al.}(2002)\citenamefont {Huo},
  \citenamefont {Kuwai}, \citenamefont {Mizushima}, \citenamefont {Isikawa},\
  and\ \citenamefont {Sakurai}}]{Huo2002Thermoelectric}%
  \BibitemOpen
  \bibfield  {author} {\bibinfo {author} {\bibfnamefont {D.}~\bibnamefont
  {Huo}}, \bibinfo {author} {\bibfnamefont {T.}~\bibnamefont {Kuwai}}, \bibinfo
  {author} {\bibfnamefont {T.}~\bibnamefont {Mizushima}}, \bibinfo {author}
  {\bibfnamefont {Y.}~\bibnamefont {Isikawa}}, \ and\ \bibinfo {author}
  {\bibfnamefont {J.}~\bibnamefont {Sakurai}},\ }\href {\doibase
  http://dx.doi.org/10.1016/S0921-4526(01)01092-4} {\bibfield  {journal}
  {\bibinfo  {journal} {Physica B}\ }\textbf {\bibinfo {volume} {312 –
  313}},\ \bibinfo {pages} {232 } (\bibinfo {year} {2002})}\BibitemShut
  {NoStop}%
\bibitem [{\citenamefont {Oyamada}\ \emph {et~al.}(2008)\citenamefont
  {Oyamada}, \citenamefont {Maegawa}, \citenamefont {Nishiyama}, \citenamefont
  {Kitazawa},\ and\ \citenamefont {Isikawa}}]{Oyamada2008Ordering}%
  \BibitemOpen
  \bibfield  {author} {\bibinfo {author} {\bibfnamefont {A.}~\bibnamefont
  {Oyamada}}, \bibinfo {author} {\bibfnamefont {S.}~\bibnamefont {Maegawa}},
  \bibinfo {author} {\bibfnamefont {M.}~\bibnamefont {Nishiyama}}, \bibinfo
  {author} {\bibfnamefont {H.}~\bibnamefont {Kitazawa}}, \ and\ \bibinfo
  {author} {\bibfnamefont {Y.}~\bibnamefont {Isikawa}},\ }\href {\doibase
  10.1103/PhysRevB.77.064432} {\bibfield  {journal} {\bibinfo  {journal} {Phys.
  Rev. B}\ }\textbf {\bibinfo {volume} {77}},\ \bibinfo {pages} {064432}
  (\bibinfo {year} {2008})}\BibitemShut {NoStop}%
\bibitem [{\citenamefont {N\'{u}\~{n}ez-Regueiro}\ \emph
  {et~al.}(1997)\citenamefont {N\'{u}\~{n}ez-Regueiro}, \citenamefont
  {Lacroix},\ and\ \citenamefont {Canals}}]{Nunez-Regueiro1997Magnetic}%
  \BibitemOpen
  \bibfield  {author} {\bibinfo {author} {\bibfnamefont {M.~D.}\ \bibnamefont
  {N\'{u}\~{n}ez-Regueiro}}, \bibinfo {author} {\bibfnamefont {C.}~\bibnamefont
  {Lacroix}}, \ and\ \bibinfo {author} {\bibfnamefont {B.}~\bibnamefont
  {Canals}},\ }\href@noop {} {\bibfield  {journal} {\bibinfo  {journal}
  {Physica C}\ }\textbf {\bibinfo {volume} {282}},\ \bibinfo {pages} {1885}
  (\bibinfo {year} {1997})}\BibitemShut {NoStop}%
\bibitem [{\citenamefont {Fritsch}\ \emph {et~al.}(2016)\citenamefont
  {Fritsch}, \citenamefont {Lucas}, \citenamefont {Huesges}, \citenamefont
  {Sakai}, \citenamefont {Kittler}, \citenamefont {Taubenheim}, \citenamefont
  {Woitschach}, \citenamefont {Pedersen}, \citenamefont {Grube}, \citenamefont
  {Schmidt}, \citenamefont {Gegenwart}, \citenamefont {Stockert},\ and\
  \citenamefont {v.~L\"{o}hneysen}}]{Fritsch2016CePdAl}%
  \BibitemOpen
  \bibfield  {author} {\bibinfo {author} {\bibfnamefont {V.}~\bibnamefont
  {Fritsch}}, \bibinfo {author} {\bibfnamefont {S.}~\bibnamefont {Lucas}},
  \bibinfo {author} {\bibfnamefont {Z.}~\bibnamefont {Huesges}}, \bibinfo
  {author} {\bibfnamefont {A.}~\bibnamefont {Sakai}}, \bibinfo {author}
  {\bibfnamefont {W.}~\bibnamefont {Kittler}}, \bibinfo {author} {\bibfnamefont
  {C.}~\bibnamefont {Taubenheim}}, \bibinfo {author} {\bibfnamefont
  {S.}~\bibnamefont {Woitschach}}, \bibinfo {author} {\bibfnamefont
  {B.}~\bibnamefont {Pedersen}}, \bibinfo {author} {\bibfnamefont
  {K.}~\bibnamefont {Grube}}, \bibinfo {author} {\bibfnamefont
  {B.}~\bibnamefont {Schmidt}}, \bibinfo {author} {\bibfnamefont
  {P.}~\bibnamefont {Gegenwart}}, \bibinfo {author} {\bibfnamefont
  {O.}~\bibnamefont {Stockert}}, \ and\ \bibinfo {author} {\bibfnamefont
  {H.}~\bibnamefont {v.~L\"{o}hneysen}},\ }\href {arXiv:1609.01551
  [cond-mat.str-el]} {\bibfield  {journal} {\bibinfo  {journal}
  {arXiv:1609.01551 [cond-mat.str-el]}\ } (\bibinfo {year} {2016})}\BibitemShut
  {NoStop}%
\bibitem [{\citenamefont {Tokiwa}\ \emph {et~al.}(2009)\citenamefont {Tokiwa},
  \citenamefont {Radu}, \citenamefont {Geibel}, \citenamefont {Steglich},\ and\
  \citenamefont {Gegenwart}}]{Tokiwa2009Divergence}%
  \BibitemOpen
  \bibfield  {author} {\bibinfo {author} {\bibfnamefont {Y.}~\bibnamefont
  {Tokiwa}}, \bibinfo {author} {\bibfnamefont {T.}~\bibnamefont {Radu}},
  \bibinfo {author} {\bibfnamefont {C.}~\bibnamefont {Geibel}}, \bibinfo
  {author} {\bibfnamefont {F.}~\bibnamefont {Steglich}}, \ and\ \bibinfo
  {author} {\bibfnamefont {P.}~\bibnamefont {Gegenwart}},\ }\href@noop {}
  {\bibfield  {journal} {\bibinfo  {journal} {Phys. Rev. Lett.}\ }\textbf
  {\bibinfo {volume} {102}},\ \bibinfo {pages} {066401} (\bibinfo {year}
  {2009})}\BibitemShut {NoStop}%
\bibitem [{\citenamefont {Tokiwa}\ and\ \citenamefont
  {Gegenwart}(2011)}]{Tokiwa2011High}%
  \BibitemOpen
  \bibfield  {author} {\bibinfo {author} {\bibfnamefont {Y.}~\bibnamefont
  {Tokiwa}}\ and\ \bibinfo {author} {\bibfnamefont {P.}~\bibnamefont
  {Gegenwart}},\ }\href@noop {} {\bibfield  {journal} {\bibinfo  {journal}
  {Rev. Sci. Inst.}\ }\textbf {\bibinfo {volume} {82}},\ \bibinfo {pages}
  {013905} (\bibinfo {year} {2011})}\BibitemShut {NoStop}%
\bibitem [{\citenamefont {Zhu}\ \emph {et~al.}(2003)\citenamefont {Zhu},
  \citenamefont {Garst}, \citenamefont {Rosch},\ and\ \citenamefont
  {Si}}]{Zhu2003Universally}%
  \BibitemOpen
  \bibfield  {author} {\bibinfo {author} {\bibfnamefont {L.}~\bibnamefont
  {Zhu}}, \bibinfo {author} {\bibfnamefont {M.}~\bibnamefont {Garst}}, \bibinfo
  {author} {\bibfnamefont {A.}~\bibnamefont {Rosch}}, \ and\ \bibinfo {author}
  {\bibfnamefont {Q.}~\bibnamefont {Si}},\ }\href {\doibase
  10.1103/PhysRevLett.91.066404} {\bibfield  {journal} {\bibinfo  {journal}
  {Phys. Rev. Lett.}\ }\textbf {\bibinfo {volume} {91}},\ \bibinfo {pages}
  {066404} (\bibinfo {year} {2003})}\BibitemShut {NoStop}%
\bibitem [{\citenamefont {Garst}\ and\ \citenamefont
  {Rosch}(2005)}]{Garst2005Sign}%
  \BibitemOpen
  \bibfield  {author} {\bibinfo {author} {\bibfnamefont {M.}~\bibnamefont
  {Garst}}\ and\ \bibinfo {author} {\bibfnamefont {A.}~\bibnamefont {Rosch}},\
  }\href {\doibase 10.1103/PhysRevB.72.205129} {\bibfield  {journal} {\bibinfo
  {journal} {Phys. Rev. B}\ }\textbf {\bibinfo {volume} {72}},\ \bibinfo
  {pages} {205129} (\bibinfo {year} {2005})}\BibitemShut {NoStop}%
\bibitem [{\citenamefont {Gribanov}\ \emph {et~al.}(2006)\citenamefont
  {Gribanov}, \citenamefont {Tursina}, \citenamefont {Murashova}, \citenamefont
  {Seropegin}, \citenamefont {Bauer}, \citenamefont {Kaldarar}, \citenamefont
  {Lackner}, \citenamefont {Michor}, \citenamefont {Royanian}, \citenamefont
  {Reissner},\ and\ \citenamefont {Rogl}}]{Gribanov2006New}%
  \BibitemOpen
  \bibfield  {author} {\bibinfo {author} {\bibfnamefont {A.}~\bibnamefont
  {Gribanov}}, \bibinfo {author} {\bibfnamefont {A.}~\bibnamefont {Tursina}},
  \bibinfo {author} {\bibfnamefont {E.}~\bibnamefont {Murashova}}, \bibinfo
  {author} {\bibfnamefont {Y.}~\bibnamefont {Seropegin}}, \bibinfo {author}
  {\bibfnamefont {E.}~\bibnamefont {Bauer}}, \bibinfo {author} {\bibfnamefont
  {H.}~\bibnamefont {Kaldarar}}, \bibinfo {author} {\bibfnamefont
  {R.}~\bibnamefont {Lackner}}, \bibinfo {author} {\bibfnamefont
  {H.}~\bibnamefont {Michor}}, \bibinfo {author} {\bibfnamefont
  {E.}~\bibnamefont {Royanian}}, \bibinfo {author} {\bibfnamefont
  {M.}~\bibnamefont {Reissner}}, \ and\ \bibinfo {author} {\bibfnamefont
  {P.}~\bibnamefont {Rogl}},\ }\href@noop {} {\bibfield  {journal} {\bibinfo
  {journal} {J. Phys.: Cond. Mat.}\ }\textbf {\bibinfo {volume} {18}},\
  \bibinfo {pages} {9593} (\bibinfo {year} {2006})}\BibitemShut {NoStop}%
\bibitem [{\citenamefont {Fritsch}\ \emph {et~al.}(2013)\citenamefont
  {Fritsch}, \citenamefont {Huang}, \citenamefont {Bagrets}, \citenamefont
  {Grube}, \citenamefont {Schumann},\ and\ \citenamefont {{v.
  L\"{o}hneysen}}}]{Fritsch2013Magnetization}%
  \BibitemOpen
  \bibfield  {author} {\bibinfo {author} {\bibfnamefont {V.}~\bibnamefont
  {Fritsch}}, \bibinfo {author} {\bibfnamefont {C.-L.}\ \bibnamefont {Huang}},
  \bibinfo {author} {\bibfnamefont {N.}~\bibnamefont {Bagrets}}, \bibinfo
  {author} {\bibfnamefont {K.}~\bibnamefont {Grube}}, \bibinfo {author}
  {\bibfnamefont {S.}~\bibnamefont {Schumann}}, \ and\ \bibinfo {author}
  {\bibfnamefont {H.}~\bibnamefont {{v. L\"{o}hneysen}}},\ }\href@noop {}
  {\bibfield  {journal} {\bibinfo  {journal} {Phys. Status Solidi B}\ }\textbf
  {\bibinfo {volume} {250}},\ \bibinfo {pages} {506 } (\bibinfo {year}
  {2013})}\BibitemShut {NoStop}%
\bibitem [{Sup()}]{Supplement}%
  \BibitemOpen
  \href@noop {} {}\bibinfo {note} {See supplemental material at [URL will be
  inserted by publisher] for the determination of the nuclear contribution to
  the specific heat and the estimation of the field dependence of the magnetic
  entropy of {\rm CePd$_{1-x}$Ni$_x$Al} single crystals at 0.2~K.}\BibitemShut
  {Stop}%
\bibitem [{\citenamefont {Garst}\ \emph {et~al.}(2008)\citenamefont {Garst},
  \citenamefont {Fritz}, \citenamefont {Rosch},\ and\ \citenamefont
  {Vojta}}]{Garst2008Dimensional}%
  \BibitemOpen
  \bibfield  {author} {\bibinfo {author} {\bibfnamefont {M.}~\bibnamefont
  {Garst}}, \bibinfo {author} {\bibfnamefont {L.}~\bibnamefont {Fritz}},
  \bibinfo {author} {\bibfnamefont {A.}~\bibnamefont {Rosch}}, \ and\ \bibinfo
  {author} {\bibfnamefont {M.}~\bibnamefont {Vojta}},\ }\href@noop {}
  {\bibfield  {journal} {\bibinfo  {journal} {Phys. Rev. B}\ }\textbf {\bibinfo
  {volume} {78}},\ \bibinfo {pages} {235118} (\bibinfo {year}
  {2008})}\BibitemShut {NoStop}%
\bibitem [{\citenamefont {v.~L\"{o}hneysen}(1996)}]{Loehneysen1996Non}%
  \BibitemOpen
  \bibfield  {author} {\bibinfo {author} {\bibfnamefont {H.}~\bibnamefont
  {v.~L\"{o}hneysen}},\ }\href@noop {} {\bibfield  {journal} {\bibinfo
  {journal} {J.~Phys.: Condens. Matter}\ }\textbf {\bibinfo {volume} {8}},\
  \bibinfo {pages} {9689 } (\bibinfo {year} {1996})}\BibitemShut {NoStop}%
\bibitem [{\citenamefont {Hamann}\ \emph {et~al.}(2013)\citenamefont {Hamann},
  \citenamefont {Stockert}, \citenamefont {Fritsch}, \citenamefont {Grube},
  \citenamefont {Schneidewind},\ and\ \citenamefont {{v.
  L\"ohneysen}}}]{Hamann2013Evolution}%
  \BibitemOpen
  \bibfield  {author} {\bibinfo {author} {\bibfnamefont {A.}~\bibnamefont
  {Hamann}}, \bibinfo {author} {\bibfnamefont {O.}~\bibnamefont {Stockert}},
  \bibinfo {author} {\bibfnamefont {V.}~\bibnamefont {Fritsch}}, \bibinfo
  {author} {\bibfnamefont {K.}~\bibnamefont {Grube}}, \bibinfo {author}
  {\bibfnamefont {A.}~\bibnamefont {Schneidewind}}, \ and\ \bibinfo {author}
  {\bibfnamefont {H.}~\bibnamefont {{v. L\"ohneysen}}},\ }\href@noop {}
  {\bibfield  {journal} {\bibinfo  {journal} {Phys. Rev. Lett.}\ }\textbf
  {\bibinfo {volume} {110}},\ \bibinfo {pages} {096404} (\bibinfo {year}
  {2013})}\BibitemShut {NoStop}%
\bibitem [{\citenamefont {Stockert}\ \emph {et~al.}(1998)\citenamefont
  {Stockert}, \citenamefont {{v. L\"{o}hneysen}}, \citenamefont {Rosch},
  \citenamefont {Pyka},\ and\ \citenamefont {Loewenhaupt}}]{Stockert1998Two}%
  \BibitemOpen
  \bibfield  {author} {\bibinfo {author} {\bibfnamefont {O.}~\bibnamefont
  {Stockert}}, \bibinfo {author} {\bibfnamefont {H.}~\bibnamefont {{v.
  L\"{o}hneysen}}}, \bibinfo {author} {\bibfnamefont {A.}~\bibnamefont
  {Rosch}}, \bibinfo {author} {\bibfnamefont {N.}~\bibnamefont {Pyka}}, \ and\
  \bibinfo {author} {\bibfnamefont {M.}~\bibnamefont {Loewenhaupt}},\
  }\href@noop {} {\bibfield  {journal} {\bibinfo  {journal} {Phys. Rev. Lett.}\
  }\textbf {\bibinfo {volume} {{80}}},\ \bibinfo {pages} {5627} (\bibinfo
  {year} {1998})}\BibitemShut {NoStop}%
\bibitem [{\citenamefont {L\"ohneysen}\ \emph {et~al.}(2001)\citenamefont
  {L\"ohneysen}, \citenamefont {Pfleiderer}, \citenamefont {Pietrus},
  \citenamefont {Stockert},\ and\ \citenamefont
  {Will}}]{Loehneysen2001Pressure}%
  \BibitemOpen
  \bibfield  {author} {\bibinfo {author} {\bibfnamefont {H.~v.}\ \bibnamefont
  {L\"ohneysen}}, \bibinfo {author} {\bibfnamefont {C.}~\bibnamefont
  {Pfleiderer}}, \bibinfo {author} {\bibfnamefont {T.}~\bibnamefont {Pietrus}},
  \bibinfo {author} {\bibfnamefont {O.}~\bibnamefont {Stockert}}, \ and\
  \bibinfo {author} {\bibfnamefont {B.}~\bibnamefont {Will}},\ }\href {\doibase
  10.1103/PhysRevB.63.134411} {\bibfield  {journal} {\bibinfo  {journal} {Phys.
  Rev. B}\ }\textbf {\bibinfo {volume} {63}},\ \bibinfo {pages} {134411}
  (\bibinfo {year} {2001})}\BibitemShut {NoStop}%
\bibitem [{\citenamefont {Stockert}\ \emph {et~al.}(2007)\citenamefont
  {Stockert}, \citenamefont {Enderle},\ and\ \citenamefont
  {L\"{o}hneysen}}]{Stockert2007Magnetic}%
  \BibitemOpen
  \bibfield  {author} {\bibinfo {author} {\bibfnamefont {O.}~\bibnamefont
  {Stockert}}, \bibinfo {author} {\bibfnamefont {M.}~\bibnamefont {Enderle}}, \
  and\ \bibinfo {author} {\bibfnamefont {H.}~\bibnamefont {L\"{o}hneysen}},\
  }\href@noop {} {\bibfield  {journal} {\bibinfo  {journal} {Phys. Rev. Lett.}\
  }\textbf {\bibinfo {volume} {99}},\ \bibinfo {pages} {237203} (\bibinfo
  {year} {2007})}\BibitemShut {NoStop}%
\end{thebibliography}
\end{document}